\documentstyle[12pt,epsf]{article}

\begin{document}

\title{Small Window Overlaps Are Effective Probes
of Replica Symmetry Breaking in 3D Spin Glasses}

\author{ Enzo Marinari$^{(a)}$, 
Giorgio Parisi$^{(b)}$ , Federico Ricci-Tersenghi$^{(b)}$ \\
and\\
Juan J. Ruiz-Lorenzo$^{(b,c)}$\\[0.5em]
$^{(a)}$
{\small Dipartimento di Fisica and INFN, Universit\`a di Cagliari}\\
{\small \ \  Via Ospedale 72, 09100 Cagliari (Italy)}\\[0.3em]
{\small \tt marinari@ca.infn.it}\\[0.5em]
$^{(b)}$
{\small Dipartimento di Fisica and INFN, Universit\`a di Roma}
{\small {\em La Sapienza} }\\
{\small \ \  P. A. Moro 2, 00185 Roma (Italy)}\\[0.3em]
$^{(c)}$
{\small Departamento de F\'{\i}sica Te\'orica, Universidad Complutense de
Madrid} \\
{\small \ \  Ciudad Universitaria s/n, 28040 Madrid (Spain)}\\[0.3em]
{\small \tt giorgio.parisi@roma1.infn.it}\\[0.3em]
{\small \tt riccife@chimera.roma1.infn.it}\\[0.3em]
{\small \tt ruiz@lattice.fis.ucm.es}\\[0.5em]
}

\date{March 22, 1998}

\maketitle

\begin{abstract}
We compute numerically small window overlaps in the three dimensional
Edwards Anderson spin glass. We show that they behave in the way
implied by the Replica Symmetry Breaking Ansatz, that they do not
qualitatively differ from the full volume overlap and do not tend to a
trivial function when increasing the lattice volume. On the contrary
we show they are affected by small finite volume effects, and are
interesting tools for the study of the features of the spin glass
phase.
\end{abstract}   

\vfill

\begin{flushright}
cond-mat/9804017
\end{flushright}

\thispagestyle{empty}

\newpage

In this note we will try to give an unambiguous answer to an important
question, concerning overlaps in spin glasses.  On one side recent
numerical simulations \cite{MAPARURI}-\cite{DINAMI} are making clear
that finite dimensional spin glasses behave in a way very reminiscent
of the Replica Symmetry Breaking (RSB) solution \cite{BREAKING,MEPAVI}
of the mean field Sherrington-Kirkpatrick model (SK) \cite{SK}.  Also
recent experimental results seem to hint that RSB feature can be
detected in real materials \cite{ORBACH}.  On the other side there
have been progresses based on rigorous and heuristic results: the
validity of the RSB solution of the SK model is supported (but not yet
proven) by the work of \cite{GUERRA,AIZCON,PARISIX}, while potential
problems in applying RSB ideas to finite dimensional spin glasses have
been stressed in \cite{NS} (but see \cite{NSREP} for ideas pointing to
the opposite direction).

It is useful, for making the issues raised in \cite{NS,NSREP} precise,
to distinguish among the full volume overlap and the small window
overlap.  In the standard case the (full volume) overlap is computed
among all the spins of two configurations of the system (under the
same realization of the quenched disorder), typically with periodic
boundary conditions.  The small window overlap is defined on a box of
size much smaller than the volume of the system.

The first kind of overlap plays an important role in RSB theory, and
it is the one that is usually measured in numerical simulations.  Only
out-of-equilibrium, dynamical measurements of the second kind of
overlap have been reported \cite{MAPARURI,BOOK}, and they are
consistent with a RSB behavior of the finite dimensional system.

The small window overlap plays an important role in trying a rigorous
analysis of the behavior of the system.  Interfaces can make the
probability distribution of the order parameter look non-trivial even
in a situation where there is no spin glass ordering, but small window
overlaps detect the difference: if there are only two equilibrium
states (related by a global flip of all the spins) the probability
distribution of the overlap in a (large) window much smaller than the
(large) size of the system will be the sum of two delta functions at
$q=\pm q_{\rm EA}$, where $q_{\rm EA}$ is the Edward-Anderson order
parameter \cite{MEPAVI}.

Let us quote from the second of references \cite{NS}: {\em [\ldots]
Essentially all the simulations of which we are aware compute the
overlap distribution in the {\em entire} box. [\ldots] we suspect that
the overlaps computed over the entire box are observing domain wall
effects arising solely from the imposed boundary conditions rather
than revealing spin glass ordering. [\ldots] In other words, if
overlap computations were measured in ``small'' windows far from any
boundary, one should find only a pair of $\delta$-functions. One way
to test this would be to fix a region at the origin, and do successive
overlap computations in that fixed region for increasingly larger
boxes with imposed periodic boundary conditions; as the boundaries
move farther away, the overlap distribution within the fixed region
should tend toward a pair of $\delta$-function.}

\begin{figure}
\begin{center}
\addvspace{1 cm}
\leavevmode
\epsfysize=250pt
\epsffile{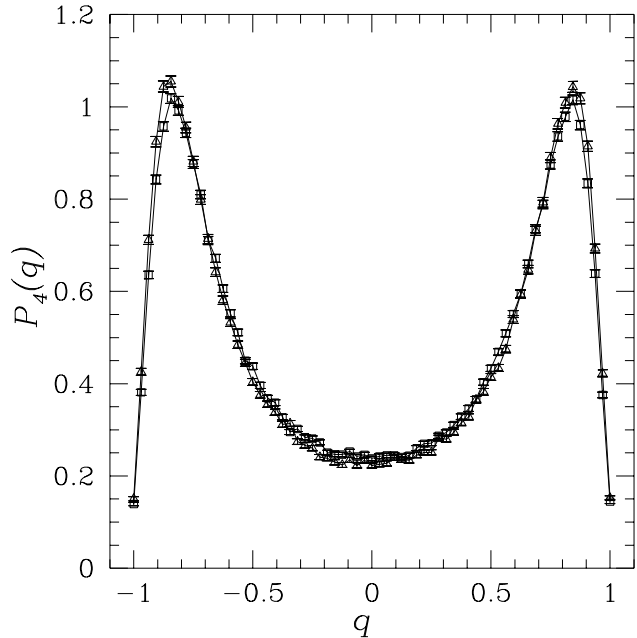}
\end{center}
\caption{$P_4(q)$: triangles for $L=8$ and squares for
$L=12$. $T=0.7$.}
\label{F-THATISIT}
\end{figure}

This is exactly what we have done, finding numerical evidence that
what happens is not what is described from the point of view we have
just quoted.  To start from the end, we show in figure
(\ref{F-THATISIT}) (that we will discuss in more detail in the rest of
the paper) the probability distribution in a block of size $B=4$
around the origin for two different lattice sizes, $L=8$ and $L=12$.
Here $T$ is lower than $T_{c}$, and the systems are at thermal
equilibrium (these are static measurements) thanks to the tempering
Monte Carlo approach \cite{TEMPERING}.  The two probability
distributions are non-trivial, and they do not have any substantial
dependence over $L$.  In no way they are approaching $\delta$-function
when the lattice volume is increasing, but they have the typical shape
of RSB probability distributions.  On the contrary, it seems (as one
would maybe expect when reasoning according to an usual point of view)
that they have smaller finite size effects than the full volume
overlap probability distribution, that is feeling more the use of
periodic boundary conditions.  We suggest indeed that small window
overlaps could turn out to be a precious tool for the numerical study
of RSB like phases: that would surely be a pleasant remainder of the
present disagreements about the behavior of finite dimensional spin
glass systems \cite{NS,NSREP}.

We have simulated the three dimensional Edward-Anderson spin glass
defined on a cubic box of size $L$ with periodic boundary
conditions. The quenched couplings $J_{ij}$ have a Gaussian
distribution with zero mean and unit variance. The Hamiltonian of the
model is

\begin{equation}
{\cal H} \equiv -\sum_{<ij>} \sigma_i J_{ij} \sigma_j\ ,
\end{equation}
where the sum runs over couples of first neighboring sites.  In the
following we will also denote by $\sigma(x,y,z)$ the spin at the point
$(x,y,z)$. We have simulated two real replicas ($\sigma$ and $\tau$)
in the same realization of the quenched disorder.

We define the overlaps (that we denote as $B$-overlaps) on a finite
cubic window (of linear size $B$), that is part of the lattice of size
$L$

\begin{equation}
q_B \equiv \frac{1}{B^3} \sum_{x=0}^{B-1} \sum_{y=0}^{B-1}
\sum_{z=0}^{B-1} \sigma(x,y,z) \tau(x,y,z) \; .
\label{B-overlap}
\end{equation}
We also define $Q_B\equiv B^3 q_B$.  We will denote by $P_B(q)$ the 
probability distribution of $q_B$.  When $B=L$ one recovers the 
standard overlap.  For every couple of spin configurations we have 
measured only the $B$-overlap related to a single origin: in 
principle one could average among all the $B$-overlaps (with a fixed 
value of $B$) centered around different sites.

We have used a $L=8$ and a $L=12$ lattice, and we have measured the
$B$-overlaps for $B=2,3,4,5$ and $6$.  We have used the parallel
tempering Monte Carlo method \cite{TEMPERING,BOOK}.  We have used a
set of $13$ temperatures, from $T=1.3$ to $T=0.7$ with a step of
$0.05$.  All the figures we will present here show data at $T=0.7$,
the lower temperature we have studied.  We have used the APE-100
parallel computer \cite{APE}, and we have simulated $2048$ samples.
The acceptance factor for the $\beta$ swap of the tempering update has
always been in the range $0.2-0.5$.  The parallel tempering $\beta$
swap has been used from the start of the thermal run.

\begin{figure}
\begin{center}
\addvspace{1 cm}
\leavevmode
\epsfysize=250pt
\epsffile{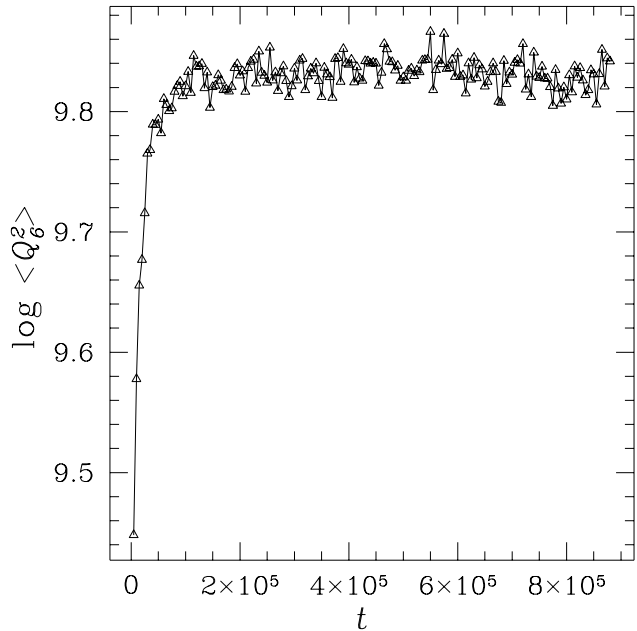}
\end{center}
\caption{$\log \overline{\langle Q_6^2 \rangle}$ versus the Monte
Carlo time $t$. $L=12$ and $T=0.7$}
\label{fig:terma}
\end{figure}

We have used all the approaches to check thermalization that are
described, for example, in \cite{MAPARU}, and we are sure of a good
thermalization of our samples.  In order to give a hint to the reader
about the situation we show in figure (\ref{fig:terma}) $\log
\overline{\langle Q_6^2 \rangle}$ at $T=0.7$ on the $L=12$ lattice
versus the Monte Carlo time.  We need $Q^{2}(t)$ to have reached a
{\em plateau} as a minimal test of thermalization.  We have chosen as
thermalization time $t_{\rm eq}=150000$.  We have redone all the
analysis shown here with a larger thermalization time, $t_{\rm
eq}=300000$, and our data do not change within the statistical error.
We have used in the analysis of the $L=8$ run $t_{\rm eq}=150000$. The
total length of the $L=8$ and $L=12$ runs was of the order of nine 
hundred thousand Monte Carlo sweeps.

\begin{figure}
\begin{center}
\addvspace{1 cm}
\leavevmode
\epsfysize=250pt
\epsffile{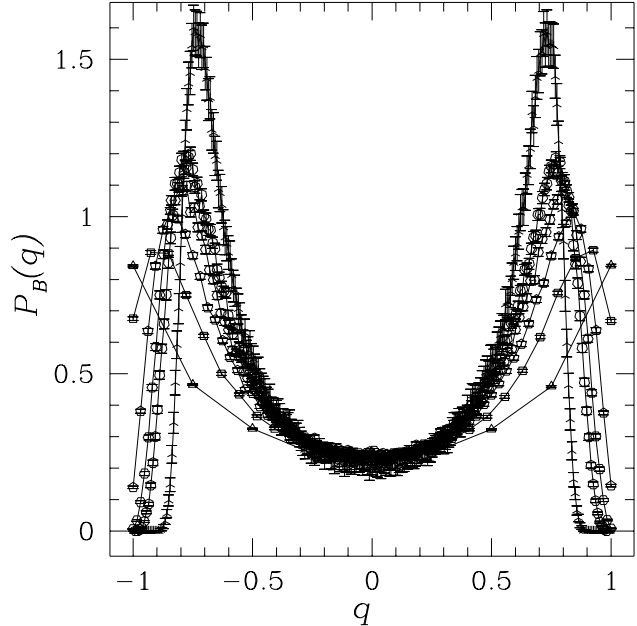}
\end{center}
\caption{$P_B(q)$ for $T=0.7$ and $L=12$ and $B=2$ (triangles), $3$
(squares), $4$ (pentagons), $5$ (hexagons), $6$ (heptagons) and $12$
(three line stars). $B$ is increasing for higher curves.}
\label{fig:pq1}
\end{figure}

Another strong thermalization test is to obtain a symmetric
probability distribution $P_B(q)$ for the $B$-overlaps under the
transformation $q\leftrightarrow -q$ \cite{TEMPERING}.  We show the
different window probability distributions at $T=0.7$ on the $L=12$
lattice in figure (\ref{fig:pq1}): all of them are fully symmetric
under the transformation $q\leftrightarrow -q$.

We have compared our window overlaps to the full volume overlap
distributions computed in \cite{MAPARU}.  These results were based on
$2048$ samples.  In that case for each sample we had run $10^6$
Metropolis steps without $\beta$ swap just to initialize the system,
followed by $10^6$ thermalization steps with Parallel Tempering and by
the real thermal run of $2\times 10^6$ parallel tempering steps, where
we measured the relevant quantities (for detail about the
thermalization of the system we refer the interested reader to
reference \cite{MAPARU}, where the issue was discussed in detail).
The $P_{12}(q)$ computed in \cite{MAPARU} appears in figure
(\ref{fig:pq1}).

Our first comments are about figure (\ref{F-THATISIT}), that makes 
clear that window overlap do not have a trivial behavior (i.e.  two 
$\delta$-functions at $\pm \overline{q}$) when the lattice volume 
increases.  The probability distribution of a block of size $4$ has 
the typical RSB shape, and basically does not change when increasing 
the lattice volume from $L=8$ to $L=12$.  For this comparison we have 
chosen $B$ as a compromise among wanting a large window, but wanting 
it still much smaller than the lattice volume.  This behavior also 
denounces that block overlaps are a very good estimator of RSB like 
effects, and they will probably play an important role in numerical 
simulations of spin glasses.

When measuring $P_B(q)$ in a finite volume simulation there are two 
different sources of finite size effects, the finiteness of the 
lattice ($L$ size) and the finiteness of the block used for the 
measurement ($B$ size).  We have found that the major changes in 
$P_B(q)$ appear when increasing the block size $B$, as shown in figure 
(\ref{fig:pq1}).  This effect is related the usual $L$-dependence of 
the full volume overlap probability distribution (see for example 
figure (6) of \cite{MAPARU}).  The great advantage of using blocks of 
fixed size (much smaller than the lattice size) is that, in this case, 
$P_B(q)$ have a very small dependence on $L$, so that we can assume 
that their shape is very similar to the one they would have in a 
$L=\infty$ lattice and we can focus our attention on their 
$B$-dependence.

In figure (\ref{fig:pq1}) we show the $L=12$, $T=0.7$ probability
distribution of the overlap $q_B$, for $B=2,3,4,5,6$ and $12$. The
shape of $P_B(q)$ changes only smoothly with the window size. The
window overlap distributions have the same qualitative behavior of the
full volume distribution, contradicting the expectations of \cite{NS}
and strongly supporting the presence of a RSB like behavior.

\begin{figure}
\begin{center}
\addvspace{1 cm}
\leavevmode
\epsfysize=250pt
\epsffile{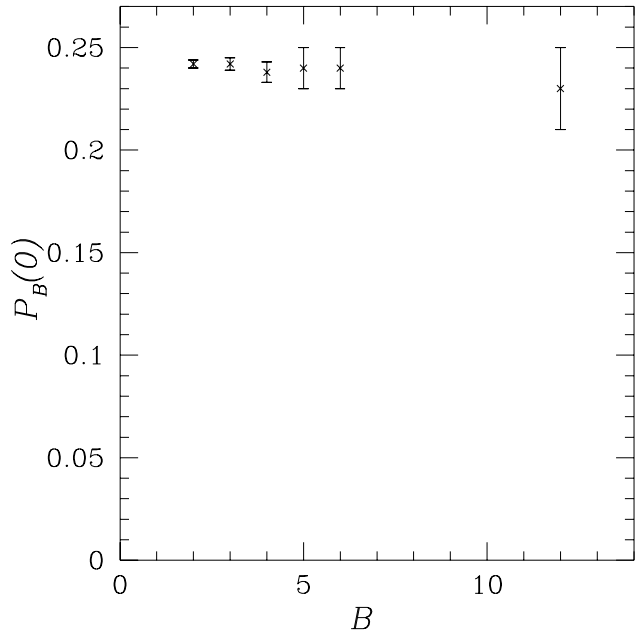}
\end{center}
\caption{$P_B(0)$ versus the block size $B$ for $L=12$, $T=0.7$.}
\label{fig:pq0}
\end{figure}

The value of $P_{B}(q_{B}\simeq 0)$ has only a very small dependence
on $B$.  In the RSB point of view it is a crucial quantity, since it
gives the probability of finding two equilibrium configurations with
very small overlap.  We plot the values of $P_B(0)$ for different $B$
values in figure (\ref{fig:pq0}). In the scenario of \cite{NS} this
number should asymptotically go to zero for $B\ll L$, while, if any,
we are observing the opposite phenomenon, i.e. a small enhancement,
due to the finiteness of the block, at small window sizes $B$.

\begin{figure}
\begin{center}
\addvspace{1 cm}
\leavevmode
\epsfysize=250pt
\epsffile{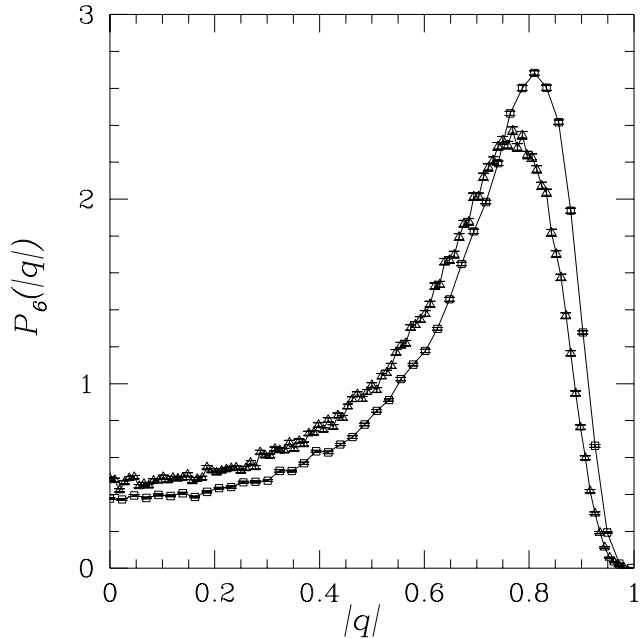}
\end{center}
\caption{$P_6(|q|)$ for $L=12$ (triangles) and for $L=6$
(squares). $T=0.7$.  The error bars are comparable to the symbol
width.}
\label{fig:pq2}
\end{figure}

Finally in figure (\ref{fig:pq2}) we show two probability 
distributions: the first one is the probability distribution computed 
in a lattice $L=6$ with $B=6$ (i.e.  it is the full volume overlap 
probability distribution on a $L=6$ lattice), while the second one is 
computed on a lattice with $L=12$ and $B=6$.  From figure 
(\ref{fig:pq2}) is clear that the shape of the two probability 
distributions is the same.  Selecting small window overlap instead 
than full volume overlaps does not imply any dramatic quantitative 
change.  We note three small effects.  The first one is that 
the overlap where the probability distribution presents the maximum is 
(slightly) lower for $P_6(q)$ with $L=12$ than for $P_6(q)$ with 
$L=6$.  The second one is that the peak of $P_6(q)$ with $L=12$ is 
lower than the one of $P_6(q)$ with $L=6$.  The third one is that the 
value of $P_6(0)$ (on the lattice with $L=12$) is slightly greater 
than the value of $P_6(0)$ (on the lattice with $L=6$).  These last 
two effects are in contradiction with the predictions of reference 
\cite{NS}: if $P_6(q)$ measured on an infinite lattice ($L = \infty$) 
was really a delta function, the value of $P_6(0)$ should decrease 
when increasing the lattice size, and the height of the peak should 
increase, with a behavior opposite to the one observed.

The results we have shown here are quite clear, and they support again
the idea that the RSB picture describes accurately the low temperature
phase of the three dimensional Ising spin glass.  We will try in a
following work \cite{NSREP} to understand better from a theoretical
point of view why the scenario proposed in \cite{NS} does not seem to
apply.

J. J. Ruiz-Lorenzo has been supported by an EC HMC (ERBFMBICT950429) grant.


\newpage

\begin{thebibliography}{99}

\bibitem{MAPARURI} 
  E. Marinari, G. Parisi, J. J. Ruiz-Lorenzo and F. Ritort, 
  Phys. Rev. Lett {\bf 76}, 843 (1996).
  
\bibitem{KAWYOU}
  N. Kawashima and A. P. Young,
  Phys. Rev. B. {\bf 53}, R484 (1996).

\bibitem{BERJAN}
  B. A. Berg and W. Janke,
  {\em Multi-Overlap Simulations of the $3d$ Edwards-Anderson Ising 
Spin Glass},
  cond-mat/9712320 (December 1997).

\bibitem{MAPARU}
  E. Marinari, G. Parisi and J.J. Ruiz-Lorenzo,
  {\em On the Phase Structure of the 3D Edwards-Anderson Spin Glass}, 
  cond-mat/9802211 (February 1998).

\bibitem{BOOK} 
E. Marinari, G. Parisi and J. J. Ruiz-Lorenzo,
{\em Numerical Simulations of Spin Glass Systems}, in {\em ``Spin
Glasses and Random Fields"}, edited by A. P. Young (World Scientific,
Singapore, 1997), 130.

\bibitem{HARTMA} 
A. K. Hartmann, 
Europhys. Lett. {\bf 40}, 429 (1997).

\bibitem{INMAPARU} 
  D. I\~niguez, E. Marinari, G. Parisi and J. J. Ruiz-Lorenzo
  J. Phys. A: Math. Gen. {\bf 30}, 7337 (1997).
  
\bibitem{4DIM}
  G. Parisi, F. Ricci-Tersenghi and J. J. Ruiz-Lorenzo,
  J. Phys. A: Math. Gen.{\bf 29}, 7943 (1996).

\bibitem{MEANFIELD} 
  G. Parisi, P. Ranieri, F. Ricci-Tersenghi and J. J. Ruiz-Lorenzo, 
  {\em Mean Field Dynamical Exponents in Finite Dimensional Ising Spin 
  Glass},
  J. Phys. A: Math. Gen. 30, 7115 (1997).

\bibitem{DINAMI}
  E. Marinari, G. Parisi, F. Ricci-Tersenghi and J. J.  Ruiz-Lorenzo, 
  {\em Violation of the Fluctuation Dissipation Theorem in Finite 
  Dimensional Spin Glasses},
  J. Phys. A: Math. Gen. 31, 2611 (1998).

\bibitem{BREAKING}
G. Parisi, 
Phys. Lett. A {\bf 73}, 154 (1979); 
Phys. Rev. Lett. {\bf 43} (1979) 1754; 
J. Phys. A: Math. Gen. {\bf 13} (1980)  L115; 1101; 1887;
Phys. Rev. Lett. {\bf 50} (1983) 1946.

\bibitem{MEPAVI}
M. M\'ezard, G. Parisi and M. A. Virasoro,
{\em Spin Glass Theory and Beyond}
(World Scientific, Singapore 1987).

\bibitem{SK}
D. Sherrington and S. Kirkpatrick,
Phys. Rev. Lett. {\bf 35},1792 (1975);
Phys. Rev. B {\bf 17}, 4384 (1978).
  
\bibitem{ORBACH} 
  Y. G. Joh, R. Orbach and J. Hamman,
  Phys. Rev. Lett. {\bf 77}, 4648 (1996).

\bibitem{GUERRA} 
  F. Guerra, 
  Int. J. Mod. Phys. B {\bf 10}, 1675 (1996).
  
\bibitem{AIZCON}
  M. Aizenman and P. Contucci,
  {\em On the Stability of the Quenched State in Mean Field Spin Glass 
       Models},
  cond-mat/9712129 (December 1997).

\bibitem{PARISIX}
  G. Parisi,
  {\em On the Probabilistic Interpretation of the Replica Approach to 
       Spin Glasses},
  cond-mat/9801081 (January 1998).

\bibitem{NS}
  C. M. Newman and D. L. Stein,
  Phys. Rev. Lett. {\bf 76}, 515 (1996);
  cond-mat/9711010 and references therein.

\bibitem{NSREP} 
  G. Parisi, 
  {\em Recent Rigorous Results Support the Predictions of Spontaneously 
  Broken Replica Symmetry for Realistic Spin Glasses},
  cond-mat/9603101 (March 1996);
  E. Marinari, G. Parisi, F. Ricci-Tersenghi, J. J. Ruiz-Lorenzo and F. 
  Zuliani,
  {\em Replica Symmetry Breaking in Short Range Spin Glasses:
  A Review of the Theoretical Foundations and of the Numerical 
  Evidence},
  to be published.
  
\bibitem{TEMPERING}
  E. Marinari and G. Parisi, Europhys. Lett. {\bf 19}, 451 (1992).
  M. C. Tesi, E. Janse van Rensburg, E. Orlandini and S. G. 
  Whillington,
  J. Stat. Phys.;
  K. Hukushima and K. Nemoto,
  J. Phys. Soc. Japan {\bf 65}, 1604 (1996);
  E. Marinari, 
  {\em Optimized Monte Carlo Methods}, 
  lectures given at the 1996 Budapest Summer School on Monte Carlo 
  Methods, edited by J. Kertesz and I. Kondor (Springer-Verlag 1997),
  cond-mat/9612010.
  
\bibitem{APE} 
  C. Battista et al., 
  Int. J. High Speed Comp. {\bf 5}, 637 (1993).

\end{thebibliography}
\end{document}